**Linear and nonlinear optical properties in spherical quantum dots: Inversely quadratic Hellmann potential**

L. Máthé[a,b*], C. P. Onyenegecha[c], A.-A. Farcaș[a,b], L.-M. Pioraș-Țimbolmaș[a,b], M. Solaimani[d], H. Hassanabadi[e]

[a]National Institute for Research and Development of Isotopic and Molecular Technologies, 67-103 Donat, 400293 Cluj-Napoca, Romania

[b]Faculty of Physics, Babeș-Bolyai University, 1 Kogălniceanu, 400084 Cluj-Napoca, Romania

[c]Faculty of Physical Sciences, Federal University of Technology Owerri, P.M.B 1526 Owerri Nigeria

[d] Department of Physics, Faculty of Mechanical Engineering, Qom University of Technology, 3718146645, Qom, Iran

[e]Faculty of Physics, Shahrood University of Technology, P. O. Box: 3619995161-316, Shahrood, Iran

*Corresponding author. E-mail address: levente.mathe@itim-cj.ro (L. Máthé)

ABSTRACT

The aims of the reported work are to provide new insights into the quantum dot optical properties confined in an inverse of a quadratic Hellmann potential. The Schrödinger equation is solved using the Nikiforov-Uvarov (NU) method, in order to obtain the analytical expressions of the eigenenergies and the eigenfunctions. The linear together with the third-order nonlinear changes in absorption coefficients and refractive index are investigated using the density matrix formalism. The absorption coefficients and the refractive index changes are strong related to the structure parameters and the optical intensity.

## 1. Introduction

The semiconductor technology has caught the attention of scientific community for ages and has shown growing interest in recent time due to the possibility to design structures with different geometrical shapes and sizes by varying the physical properties that can be of particular interest for potential applications in high speed electro-optical modulators, infrared photodetectors and other semiconductor devices [1,2].

The quantum confinement effects in low dimensional semiconductor materials could induce nonlinear optical effects, essential in the development of electronic and optoelectronic devices. The linear and nonlinear optical properties of quantum wells, quantum wires and quantum dots (QDs) have been investigated both theoretically and experimentally [3-10].

The QDs unique electronic and optical properties are determined by the intersubbands optical transitions. The incident photons with energies equal to the intersubband transition energy can discover significantly changes in the dielectric constant of material modifying thus its absorption coefficient and refractive index. The dipole matrix elements associated with the intersubband optical transitions of QDs can achieve large values. The nonlinear parts of optical properties in QDs can be enhanced by increasing the value of the dipole matrix element and decreasing the energy between subbands at the same time [11-12].

The electronic and optical properties of low dimensional QDs have been investigated using various potential models such as Tietz [13], Gaussian [14], modified Gaussian [15], modified Kratzer [16], Rosen-Morse [17], ring-shaped non-spherical oscillator [18], Manning-Rosen [19], Hulthén [20], Woods-Saxon [21,22] and other potentials. However, to the best of our knowledge, the optical properties of QDs have not been studied using the inversely quadratic Hellmann (IQH) potential.

The purpose of the present work is to theoretically study the changes in both the absorption coefficients and the refractive index in spherical QDs using IQH potential. To this intent, the linear and nonlinear absorption coefficients and refractive index changes are obtained applying the density matrix formalism. Initially, the Schrödinger equation is solved with IQH potential using the NU method. Then, the reported study, focuses on the optical properties of the system.

## 2. Theory

### 2.1. Model

Within the framework of an effective mass approximation, the Schrödinger equation in spherical coordinates can be expressed for an electron confined by a spherical QD as

$$-\frac{\hbar^2}{2\mu}\left[\frac{1}{r^2}\frac{\partial}{\partial r}\left(r^2\frac{\partial}{\partial r}\right)+\frac{1}{r^2\sin\theta}\frac{\partial}{\partial \theta}\left(\sin\theta\frac{\partial}{\partial \theta}\right)+\frac{1}{r^2\sin^2\theta}\frac{\partial^2}{\partial \phi^2}\right]\psi_{nlm}(r,\theta,\phi)$$
$$+V(r)\psi_{nlm}(r,\theta,\phi)=E_{nl}\psi_{nlm}(r,\theta,\phi) \quad (1)$$

where $\mu$ is the effective mass of the electron, $\hbar$ is the reduced Planck's constant, $n$, $l$ and $m$ are the principal, orbital and magnetic quantum numbers, $V(r)$ is the confining potential and the wave function is expressed as:

$$\psi_{nlm}(r,\theta,\phi)=R_{nl}(r)Y_l^m(\theta,\phi), \quad (2)$$

where $Y_l^m(\theta,\phi)$ are the spherical wave functions. For the radial wave function $R_{nl}(r)$ the following equation is obtained:

$$\frac{d^2 R_{nl}(r)}{dr^2}+\frac{2}{r}\frac{dR_{nl}(r)}{dr}+\frac{2\mu}{\hbar^2}\left[E_{nl}-V(r)-\frac{\hbar^2 l(l+1)}{2\mu}\frac{1}{r^2}\right]R_{nl}(r)=0 \quad (3)$$

In this work, the IQH potential is considered as the confinement potential and is given by [23-25]:

$$V(r)=-\frac{A}{r}+\frac{B}{r^2}e^{-\alpha r}, \quad (4)$$

where $\alpha$ is an adjustable screening parameter which defines the dot radius $R_0=1/\alpha$. Therefore, $A$ and $B$ are connected to the potential height $V_0$, and they are considered as $A=V_0\cdot R_0$ and $B=V_0\cdot R_0^2$, respectively. For small values of $\alpha$ Eq. (4) simplifies as

$$V(r)=\frac{B}{r^2}-\frac{A+\alpha B}{r}+\frac{\alpha^2}{2}B=V_0\left[\frac{1}{2}-2\frac{R_0}{r}+\left(\frac{R_0}{r}\right)^2\right]. \quad (5)$$

Substituting thus the first expression of Eq. (5) into Eq. (3), the radial part of the Schrödinger equation changes to:

$$\frac{d^2R_{nl}(r)}{dr^2}+\frac{2}{r}\frac{dR_{nl}(r)}{dr}+\frac{2\mu}{\hbar^2}\left[E_{nl}-\left(\frac{\hbar^2l(l+1)}{2\mu}+B\right)\frac{1}{r^2}+\frac{A+\alpha B}{r}-\frac{\alpha^2}{2}B\right]R_{nl}(r)=0. \tag{6}$$

Eq. (6) reduces to the form:

$$\frac{d^2R_{nl}(r)}{dr^2}+\frac{2}{r}\frac{dR_{nl}(r)}{dr}+\frac{-C_1r^2+C_2r-C_3}{r^2}R_{nl}(r)=0, \tag{7}$$

where

$$C_1=-\frac{2\mu}{\hbar^2}\left(E_{nl}-\frac{\alpha^2}{2}B\right),$$

$$C_2=\frac{2\mu}{\hbar^2}(A+\alpha B), \tag{8}$$

$$C_3=\frac{2\mu}{\hbar^2}B+l(l+1).$$

In order to obtain the eigenvalues and eigenfunctions of Eq. (7), the NU method is used [23-24,26-30]. Starting from the following equation:

$$\frac{d^2\psi(s)}{ds^2}+\frac{\alpha_1-\alpha_2 s}{s(1-\alpha_3 s)}\frac{d\psi(s)}{ds}+\frac{-\xi_1 s^2+\xi_2 s-\xi_3}{s^2(1-\alpha_3 s)^2}\psi(s)=0. \tag{9}$$

After applying the parametric NU method as proposed by Tezcan and Sever [27], the equation has the following solution:

$$\psi(s)=N_{nl}s^{\alpha_{12}}(1-\alpha_3 s)^{-\alpha_{12}-(\alpha_{13}/\alpha_3)}P_n^{(\alpha_{10}-1,\frac{\alpha_{11}}{\alpha_3}-\alpha_{10}-1)}(1-2\alpha_3 s), \tag{10}$$

where $P_n^{(\alpha,\beta)}$ are the Jacobi polynomials. In the special case: $\alpha_3=0$, Eq. (10) reduces to

$$\psi(s)=N_{nl}s^{\alpha_{12}}e^{\alpha_{13}\cdot s}L_n^{\alpha_{10}-1}(\alpha_{11}\cdot s), \tag{11}$$

where $L_n$ represents the Laguerre polynomials. The equation for the energy is written as:

$$(\alpha_2-\alpha_3)n+\alpha_3 n^2-(2n+1)\alpha_5+(2n+1)(\sqrt{\alpha_9}+\alpha_3\sqrt{\alpha_8})+\alpha_7+2\alpha_3\alpha_8+2\sqrt{\alpha_8\alpha_9}=0, \tag{12}$$

with

$$\left.\begin{aligned}
&\alpha_4 = \frac{1}{2}(1-\alpha_1),\ \alpha_5 = \frac{1}{2}(\alpha_2 - 2\alpha_3),\ \alpha_6 = \alpha_5^2 + \xi_1,\ \alpha_7 = 2\alpha_4\alpha_5 - \xi_2, \\
&\alpha_8 = \alpha_4^2 + \xi_3,\ \alpha_9 = \alpha_3\alpha_7 + \alpha_3^2\alpha_8 + \alpha_6,\ \alpha_{10} = \alpha_1 + 2\alpha_4 + 2\sqrt{\alpha_8}, \\
&\alpha_{11} = \alpha_2 - 2\alpha_5 + 2\left(\sqrt{\alpha_9} + \alpha_3\sqrt{\alpha_8}\right),\ \alpha_{12} = \alpha_4 + \sqrt{\alpha_8},\ \alpha_{13} = \alpha_5 - \left(\sqrt{\alpha_9} + \alpha_3\sqrt{\alpha_8}\right).
\end{aligned}\right\} \quad (13)$$

Now, comparing Eq. (7) with Eq. (9), the parameters take the following values:

$$\alpha_1 = 2,\ \alpha_2 = \alpha_3 = 0\ \text{with}\ \xi_1 = C_1,\ \xi_2 = C_2,\ \xi_3 = C_3. \quad (14)$$

The values of the parametric constants are obtained from Eq. (13):

$$\left.\begin{aligned}
&\alpha_4 = -\frac{1}{2},\ \alpha_5 = 0,\ \alpha_6 = C_1,\ \alpha_7 = -C_2,\ \alpha_8 = C_3 + \frac{1}{4},\ \alpha_9 = C_1, \\
&\alpha_{10} = 1 + 2\sqrt{C_3 + \frac{1}{4}},\ \alpha_{11} = 2\sqrt{C_1},\ \alpha_{12} = -\frac{1}{2} + \sqrt{C_3 + \frac{1}{4}},\ \alpha_{13} = -\sqrt{C_1}.
\end{aligned}\right\} \quad (15)$$

Using Eqs. (12), (14) and (15), the energy levels in QD become:

$$E_{nl} = \frac{\alpha^2}{2}B - \frac{2\mu}{\hbar^2}\left(\frac{A + \alpha B}{2n + 1 + \sqrt{(2l+1)^2 + \frac{8\mu}{\hbar^2}B}}\right)^2 \quad (16)$$

which can also be expressed in the following form:

$$E_{nl} = \frac{V_0}{2} - \frac{8\mu}{\hbar^2}\left(\frac{V_0 R_0}{2n + 1 + \sqrt{(2l+1)^2 + \frac{8\mu}{\hbar^2}V_0 R_0^2}}\right)^2. \quad (17)$$

According to Eq. (11),

$$R_{nl}(r) = N_{nl}\, r^{\frac{1}{2}(\chi_l - 1)} e^{-C_{nl} r} L_n^{\chi_l}(2C_{nl} r), \quad (18)$$

where

$$\chi_l = \sqrt{(2l+1)^2 + \frac{8\mu}{\hbar^2}V_0 R_0^2}, \quad (19)$$

and

$$C_{nl} = \sqrt{\frac{2\mu}{\hbar^2}\left(\frac{V_0}{2} - E_{nl}\right)}. \tag{20}$$

The $N_{nl}$ normalization constant can be determined from the condition:

$$\int \psi_{nlm}(r,\theta,\phi) \cdot \psi^*_{nlm}(r,\theta,\phi)\,d\Omega = 1, \tag{21}$$

where $d\Omega = r^2 dr \sin\theta\, d\theta d\phi$.

*2.2. Optical properties: theoretical details*

In this section, the density matrix formalism is applied to determine the changes in absorption coefficients and refractive index associated to an optical transition between two subbands [17,19,20]. In order to explain the intersubband transition phenomena, a general quadratic-like confining potential is considered (see Fig. 1). The ground and first excited states of the quantum dot are indicated by the corresponding energy levels $E_1$ and $E_2$, respectively. The discrete energy levels represent the subbands. The intersubband transitions can occur between the ground and the first excited subband, when the incident photon energy $\hbar\omega$ equals the energy difference between subbands $\hbar\omega = E_{21} = E_2 - E_1$. In this way, the photon absorption process induces a transition of an electron from the ground state to the excited one. The absorption process is highly influenced by the photon energy and the structure parameters. The incident photons are produced by an electromagnetic radiation which is polarized along z-direction. The subband transition induces a medium polarization due to the arranging of the electric dipole moments along the direction of electromagnetic field [31]. The exciting electromagnetic field is considered as

$$E(t) = E_0 \cos(\omega t) = \tilde{E} e^{i\omega t} + \tilde{E}^* e^{-i\omega t}, \tag{22}$$

where $\tilde{E}$ and $\omega$ are the amplitude and the angular frequency of electric field [18,32]. The time evolution of the matrix elements is calculated in the case of one-electron density operator $\rho$ [4,5,17-20,32-34]:

$$\frac{\partial \rho}{\partial t} = \frac{1}{i\hbar}[H_0 - ezE(t), \rho] - \Gamma(\rho - \rho^{(0)}) \tag{23}$$

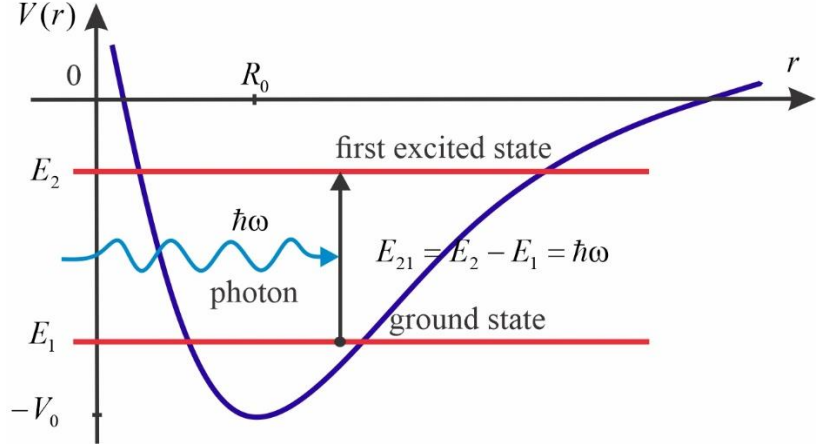

Fig. 1. Schematic diagram for energy levels of a quantum dot with an arbitrary quadratic confining potential $V(r)$ in the real space $r$. $V_0$ and $R_0$ represent the potential height and the quantum dot radius. $E_1$ and $E_2$ are the ground state and the first excited state energies, respectively. $E_{21}$ is the energy difference between the two electronic states and $\hbar\omega$ denotes the incident photon energy.

where $H_0$ is the system Hamiltonian in absence of electromagnetic field $E(t)$ with $e$ being the electronic charge, $\rho^{(0)}$ is the unperturbed density matrix operator and $\Gamma$ is a damping operator due to the collision processes [4]. $\Gamma$ is assumed to be a diagonal matrix with elements equal to the inverse of the relaxation time $\tau$ [17,19,20]. Eq. (23) can be solved using an iterative method [17,19,20,32-34]:

$$\rho(t) = \sum_{n=0}^{\infty} \rho^{(n)}(t), \qquad (24)$$

with

$$\frac{\partial \rho_{ij}^{(n+1)}}{\partial t} = \frac{1}{i\hbar}\left\{[H_0, \rho^{(n+1)}]_{ij} - i\hbar\Gamma_{ij}\rho_{ij}^{(n+1)}\right\} - \frac{1}{i\hbar}[ez, \rho^{(n)}]_{ij} E(t). \qquad (25)$$

Due to the electric field, the electronic polarization $P(t)$ can be expressed as

$$P(t) = \varepsilon_0 \chi(\omega)\tilde{E}e^{-i\omega t} + \varepsilon_0 \chi(-\omega)\tilde{E}^* e^{i\omega t} = \left(\frac{1}{V}\right) Tr(\rho M), \qquad (26)$$

where $M$ is the dipole operator, $\chi(\omega)$ is the susceptibility, $V$ is the system volume and $\varepsilon_0$ is the permittivity of vacuum. Expanding the susceptibility $\chi(\omega)$ in series and taking into account only the coefficients up to the third-order, the even orders of susceptibility vanish, due to the inversion symmetry of the system [4]. The analytical expressions of the linear $\chi^{(1)}$ and the third-order nonlinear $\chi^{(3)}$ susceptibility coefficients are derived [4,5,17-20,32]:

$$\varepsilon_0 \chi^{(1)}(\omega) = \frac{\sigma_v |M_{21}|^2}{E_{21} - \hbar\omega - i\hbar\Gamma_{12}}, \tag{27}$$

and

$$\varepsilon_0 \chi^{(3)}(\omega) = -\frac{\sigma_v |M_{21}|^2 |\tilde{E}|^2}{E_{21} - \hbar\omega - i\hbar\Gamma_{12}} \left[ \frac{4|M_{21}|^2}{(E_{21} - \hbar\omega)^2 + (\hbar\Gamma_{12})^2} - \frac{|M_{22} - M_{11}|^2}{(E_{21} - i\hbar\Gamma_{12})(E_{21} - \hbar\omega - i\hbar\Gamma_{12})} \right], \tag{28}$$

where $M_{ij} = \langle \psi_{n_i \ell_i m_i} | ez | \psi_{n_j \ell_j m_j} \rangle$ is the matrix elements of the electric dipole moment, $E_{ij} = E_i - E_j$ is the energy difference between the two electronic states, $\hbar\omega$ is the incident photon energy and $\sigma_v$ is the carrier density. The matrix elements of electric dipole moment can be calculated from

$$M_{ij} = e \int \psi^*_{n_i \ell_i m_i}(r,\theta,\phi) z \, \psi_{n_j \ell_j m_j}(r,\theta,\phi) r^2 dr \sin\theta \, d\theta \, d\phi, \tag{29}$$

where $z = r\cos\theta$ and the wave function is expressed as

$$\psi_{nlm}(r,\theta,\phi) = N_{nl} \, r^{\frac{1}{2}(\chi_l - 1)} e^{-C_{nl} r} L_n^{\chi_l}(2C_{nl} r) Y_l^m(\theta,\phi). \tag{30}$$

The matrix element of electric dipole moment is proportional with the intersubband optical transition probability amplitude between the two electronic states, such as the ground (initial) and the first excited (final) state, respectively. Due to the fact that the electric dipole moment is independent of the spin, thus the optical transitions are allowed only between the same spin states. The matrix elements of electric dipole moment are non-vanishing when the selection rules $\Delta l = \pm 1$ and $\Delta m = 0$ are satisfied [3,35].

The absorption coefficient $\alpha(\omega)$ is obtained from the susceptibility $\chi(\omega)$ via the following relation [17,19,20,32-34]:

$$\alpha(\omega) = \omega \sqrt{\frac{\mu_0}{\varepsilon}} \text{Im}[\varepsilon_0 \chi(\omega)], \tag{31}$$

where $\mu_0$ is the vacuum permeability and $\varepsilon = n_r^2 \varepsilon_0$ is the permittivity of the material with $n_r$ being the refractive index. From Eqs. (27), (28) and (31), the linear and third-order nonlinear absorption coefficients can therefore be written as

$$\alpha^{(1)}(\omega) = \omega \sqrt{\frac{\mu_0}{\varepsilon}} \frac{\sigma_v \hbar\Gamma_{12} |M_{21}|^2}{(E_{21} - \hbar\omega)^2 + (\hbar\Gamma_{12})^2}, \tag{32}$$

and

$$\alpha^{(3)}(\omega, I) = -\omega \sqrt{\frac{\mu_0}{\varepsilon}} \frac{I}{2\varepsilon_0 n_r c} \frac{\sigma_v \hbar \Gamma_{12} |M_{21}|^2}{\left[(E_{21} - \hbar\omega)^2 - (i\hbar\Gamma_{12})^2\right]^2} \times$$
$$\times \left\{ 4|M_{21}|^2 - \frac{|M_{22} - M_{11}|^2 \left[3E_{21}^2 - 4E_{21}\hbar\omega + \hbar^2(\omega^2 - \Gamma_{12}^2)\right]}{E_{21}^2 - (i\hbar\Gamma_{12})^2} \right\} \tag{33}$$

where $c$ is the speed of light in vacuum and $I = 2\varepsilon_0 n_r c |\tilde{E}|^2$ represents the incident optical intensity. Due to the spherical symmetry of the system ($M_{ii} = 0$), the third-order nonlinear absorption coefficient $\alpha^{(3)}(\omega, I)$ is negative and also is proportional to the incident optical intensity $I$. The total absorption coefficient is given by [17,19,20,32,34]

$$\alpha(\omega, I) = \alpha^{(1)}(\omega) + \alpha^{(3)}(\omega, I). \tag{34}$$

The refractive index is connected to the susceptibility as [32-34]:

$$\frac{\Delta n(\omega)}{n_r} = \text{Re}\left(\frac{\chi(\omega)}{2n_r^2}\right). \tag{35}$$

Using the Eqs. (27), (28) and (35), the linear and third-order nonlinear refractive index changes can be obtained as

$$\frac{\Delta n^{(1)}(\omega)}{n_r} = \frac{\sigma_v |M_{21}|^2}{2n_r^2 \varepsilon_0} \frac{E_{21} - \hbar\omega}{(E_{21} - \hbar\omega)^2 + (\hbar\Gamma_{12})^2}, \tag{36}$$

and

$$\frac{\Delta n^{(3)}(\omega)}{n_r} = -\frac{\sigma_v |M_{21}|^2}{4n_r^2 \varepsilon_0} \frac{\mu_0 cI}{[(E_{21} - \hbar\omega)^2 + (\hbar\Gamma_{12})^2]^2} \left\{ 4(E_{21} - \hbar\omega)|M_{21}|^2 - \frac{|M_{22} - M_{11}|^2}{(E_{21} - \hbar\omega)^2 + (\hbar\Gamma_{12})^2} \times \right. \tag{37}$$
$$\left. \times \left[(E_{21} - \hbar\omega)[E_{21}(E_{21} - \hbar\omega) - (\hbar\Gamma_{12})^2] - (\hbar\Gamma_{12})^2(2E_{21} - \hbar\omega)\right] \right\}.$$

In Eq. (37), the third-order nonlinear refractive index changes $\Delta n^{(3)}(\omega)/n_r$ has opposite sign with the sign of $\Delta n^{(1)}(\omega)/n_r$ and is proportional to the incident optical intensity $I$. Using Eqs. (36) and (37), one can express the total refractive index change as [17,19,20,32-34]

$$\frac{\Delta n(\omega)}{n_r} = \frac{\Delta n^{(1)}(\omega)}{n_r} + \frac{\Delta n^{(3)}(\omega)}{n_r}. \tag{38}$$

As a consequence, the absorption coefficients and refractive index changes given by Eqs. (32)-(34) and (36)-(38), respectively, will only depend on the matrix element $M_{21}$ of the dipole moment. When the system is irradiated by one photon with energy $\hbar\omega$, the matrix element $M_{21}$ gives the probability amplitude of one electron optical transition between the state $|1\rangle$ and $|2\rangle$ described by the wave functions $\psi_{n_1 \ell_1 m_1}$ and $\psi_{n_2 \ell_2 m_2}$, respectively.

## 3. Results and Discussions

The numerical results are computed for the GaAs QD structure. The following parameters are used: $\mu = 0.067\, m_e$, $\sigma_v = 5 \cdot 10^{22} m^{-3}$ and $n_r = 3.2$ [19,20]. Therefore, $\varepsilon_0 = 8.854 \cdot 10^{-12} F/m$ is the vacuum permittivity, $\mu_0 = 4\pi \cdot 10^{-7} H/m$ is the vacuum permeability, and $\Gamma_{12} = 1/T_{12}$ where $T_{12} = 0.14\, ps$. The values of potential height $V_0$ and dot radius $R_0$ are chosen according to the literature [14,16,36]. The quantum dot was created in a semiconductor heterostructure consisting a two dimensional electron gas, by lithographically-realized metallic gate electrodes located near the quantum dot. The energy levels and the dot size such as the dot radius can be modified by tuning the gate electrodes [37-38]. For optical properties calculations, the ground and the excited states are chosen as: $n = 0$, $l = 0$ and $n = 0$, $l = 1$. The calculations are performed in atomic units ( $\hbar = m_e = e = 1$).

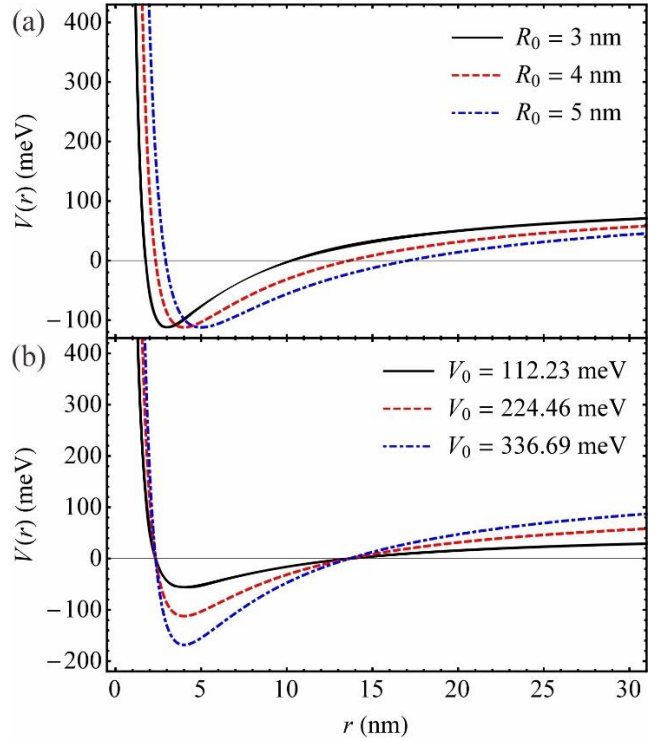

Fig. 2. (a) The confining IQH potential as a function of distance $r$ for different values of the dot radius $R_0 = 3, 4, 5\, \text{nm}$, where the potential height is set to be $V_0 = 224.46\, \text{meV}$. (b) The confining IQH potential as a function of distance $r$ for three different values of the potential height $V_0 = 112.23, 224.46, 336.69\, \text{meV}$ with dot radius $R_0 = 4\, \text{nm}$.

In Fig. 2, the confining IQH potential given by Eq. (5) is plotted against the distance $r$ for various values of the (a) dot radius $R_0$ and (b) potential height $V_0$. One can find that the potential is a continuous function which is characterized by the radius of QD $R_0$ and the height of the potential $V_0$. The IQH potential has a local minimum at the distance $R_0$, corresponding to the dot radius, because of $[\partial^2 V(r)/\partial r^2]_{r=R_0} > 0$. The depth of the potential at the minimum point is calculated as $V(r = R_0) = -V_0/2$ which can be modified by changing the height of confining potential $V_0$. It can be verified that for higher values of the distance $r$, the value of $V(r)$ becomes independent of dot radius $R_0$ and tends to a constant value, which is only determined by the potential height $V_0$, i.e., $\lim_{r \to \infty} V(r) = V_0/2$. Therefore, the potential $V(r)$ changes sign at its roots $r = (2 \pm \sqrt{2})R_0$ which can be determined from the relation $V(r) = 0$.

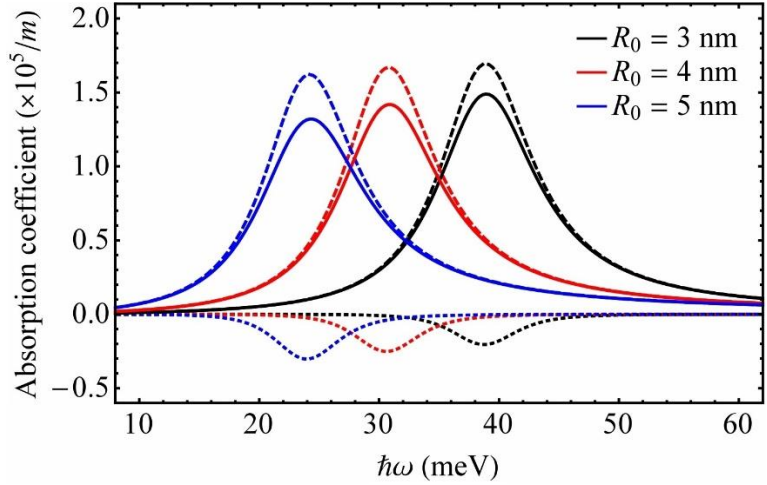

Fig. 3. The linear (dashed line) $\alpha^{(1)}(\omega)$, third-order nonlinear (dotted line) $\alpha^{(3)}(\omega)$ and total (solid line) $\alpha(\omega)$ optical absorption coefficients as a function of the incident photon energy $\hbar\omega$ for three different values of the dot radius $R_0 = 3, 4, 5\,\text{nm}$ with potential height $V_0 = 224.46\,\text{meV}$ and optical intensity $I = 8 \times 10^8\,\text{W/m}^2$.

In order to investigate the influence of $R_0$, radius of QD, on absorption coefficients, the linear, third-order nonlinear and total absorption coefficients are plotted as a function of the incident photon energy $\hbar\omega$ with optical intensity $I = 8 \times 10^8\,\text{W/m}^2$ and potential height $V_0 = 224.46\,\text{meV}$ for three different values of the QD radius in Fig. 3. When the photon energy equals the energy difference between the subbands $\hbar\omega = E_{21}$, the linear and the total absorption coefficients present a maximum, while the third-order nonlinear absorption coefficient has a minimum. It can be observed that the peaks of linear, third-order nonlinear as well as total optical absorption coefficients shift to the lower energies (red shift) with increasing the radius of the dot $R_0$. This fact is due to the decreasing energy difference between the ground and excited states ($E_{21}$) caused by the increasing of the dot radius $R_0$. As a matter of fact, the effect of radius change on the amplitude of absorption coefficients is more visible in the case of third-order nonlinear absorption coefficient. This behavior is due to the higher-order dependence of the third-order nonlinear absorption coefficient on the transition matrix element ($M_{21}$), its value increases with enhancement of the radius of QD. The enlarged matrix element leads to the third-order nonlinear absorption peaks become increased and sharper.

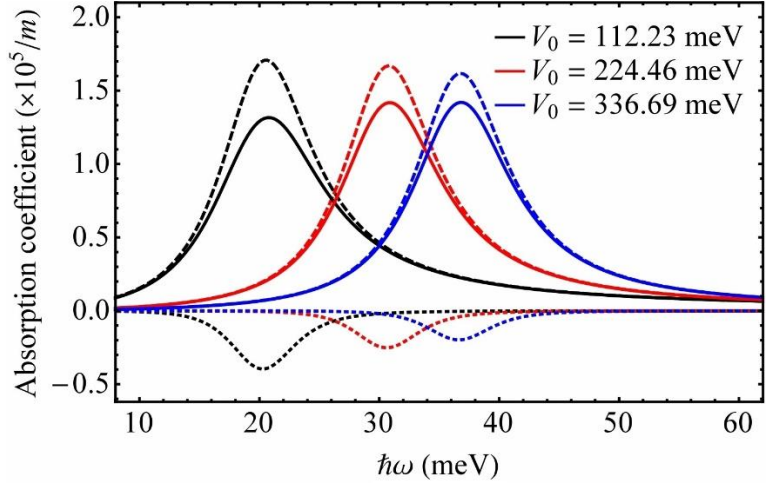

Fig. 4. The linear (dashed line) $\alpha^{(1)}(\omega)$, third-order nonlinear (dotted line) $\alpha^{(3)}(\omega)$ and total (solid line) $\alpha(\omega)$ optical absorption coefficients as a function of the incident photon energy $\hbar\omega$ for three different values of the potential height $V_0 = 112.23, 224.46, 336.69$ meV with dot radius $R_0 = 4$ nm and optical intensity $I = 8 \times 10^8$ W/m$^2$.

In Fig. 4, the linear, third-order nonlinear and total absorption coefficients are plotted as a function of the incident photon energy $\hbar\omega$ for three different barrier heights $V_0$ with QD radius $R_0 = 4$ nm and incident optical intensity $I = 8 \times 10^8$ W/m$^2$. One can find that the position of resonant peaks of absorption coefficients shift to the higher energies (blue shift) with increasing the barrier height $V_0$. This behavior of peaks can be due to the fact that the energy difference between the ground state and first excited state increases when the height of potential $V_0$ increases. The matrix element $M_{21}$ decreases when the potential height $V_0$ increases resulting in that the amplitude of third-order nonlinear absorption coefficient will decrease and become more broadened.

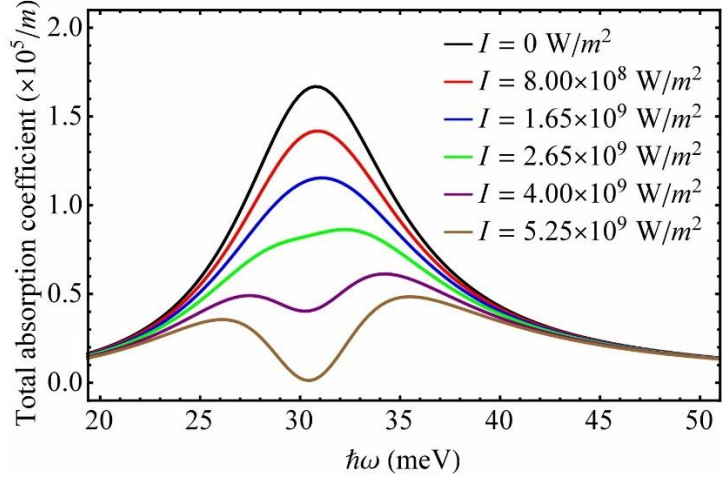

Fig. 5. The total optical absorption coefficient $\alpha(\omega)$ as a function of the incident photon energy $\hbar\omega$ for six different values of the incident optical intensity $I = 0, \ 8\times10^8, 1.65\times10^9, 2.65\times10^9, 4\times10^9, 5.25\times10^9 \ \text{W/m}^2$ with dot radius $R_0 = 4\,\text{nm}$ and potential height $V_0 = 224.46\,\text{meV}$.

Fig. 5 shows the total changes in the absorption coefficient as a function of the photon energy $\hbar\omega$ for six different incident optical intensities with the dot radius $R_0 = 4\,\text{nm}$ and potential height $V_0 = 224.46\,\text{meV}$. It can be noted that the amplitude of the total absorption coefficient is decreasing with increasing the incident photon intensity. This is due to the fact that the linear absorption coefficient is independent of *I*, while the third-order nonlinear absorption coefficient depends linearly on *I*. With increasing the intensity *I*, the total absorption coefficient reduces and begins to saturate at the critical optical intensity $I \approx 2.65\times10^9 \ \text{W/m}^2$, when the magnitude of both total and third-order nonlinear absorption coefficients become the same, i.e., $\alpha = |\alpha^{(3)}|$, at the resonant frequency $\hbar\omega = E_{21}$. As *I* increases, the third-order nonlinear term becomes dominant and results in a dip of the total absorption coefficient. The dip in the total absorption coefficient $\alpha(\omega)$ originates from the destructive interference phenomena, which take place between the one and three photon absorption processes, corresponding to $\alpha^{(1)}(\omega)$ and $\alpha^{(3)}(\omega)$, respectively. When the magnitude of linear and third-order nonlinear absorption coefficients are equal $\alpha^{(1)} = |\alpha^{(3)}|$, at the optical intensity $I \approx 5.25 \times 10^9 \, \text{W/m}^2$, the total absorption coefficient reaches the zero value. Note that further increase of *I*, leads to negative values of the total absorption coefficient $\alpha(\omega)$. The resonance peak position of total absorption coefficient is not influenced by the optical intensity.

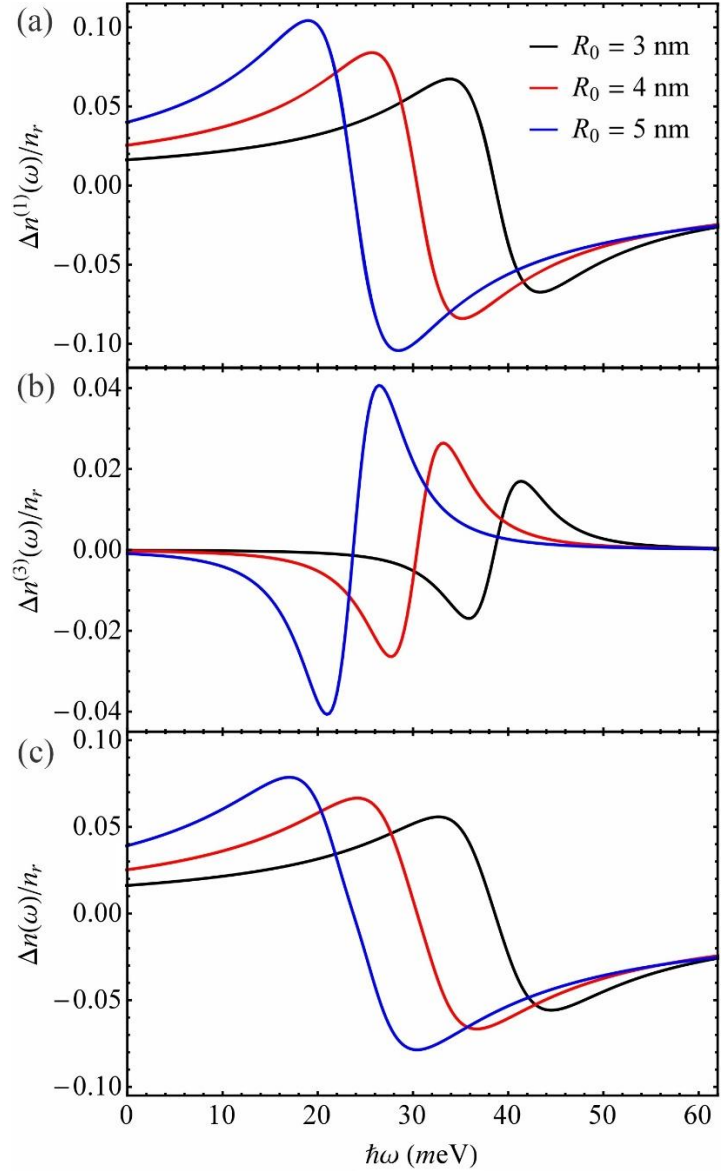

Fig. 6. The (a) linear $\Delta n^{(1)}(\omega)/n_r$, (b) third-order nonlinear $\Delta n^{(3)}(\omega)/n_r$ and (c) total $\Delta n(\omega)/n_r$ relative changes in the refractive index as a function of the incident photon energy $\hbar\omega$ for three different values of the dot radius $R_0 = 3, 4, 5 \, \text{nm}$ with potential height $V_0 = 224.46 \, \text{meV}$ and optical intensity $I = 8 \times 10^8 \, \text{W/m}^2$.

Fig. 6 illustrates the results for the linear, third-order nonlinear and total changes in refractive index as a function of the incident photon energy $\hbar\omega$ for different values of the QD radius $R_0$. The barrier height and incident optical intensity are set to be $V_0 = 224.46\,\text{meV}$ and $I = 8\times10^8\,\text{W/m}^2$, respectively. The linear, third-order nonlinear and total refractive index changes match the zero value at the photon energy $\hbar\omega = E_{21}$. It can be found that the peaks shift to the lower energies (red shift) when the radius of QD increases. This is because the energy difference between the ground state and first excited state decreases with increasing the dot radius $R_0$. The magnitude of peaks enhances when the radius of QD increases which is caused by the enlarging matrix element $M_{21}$.

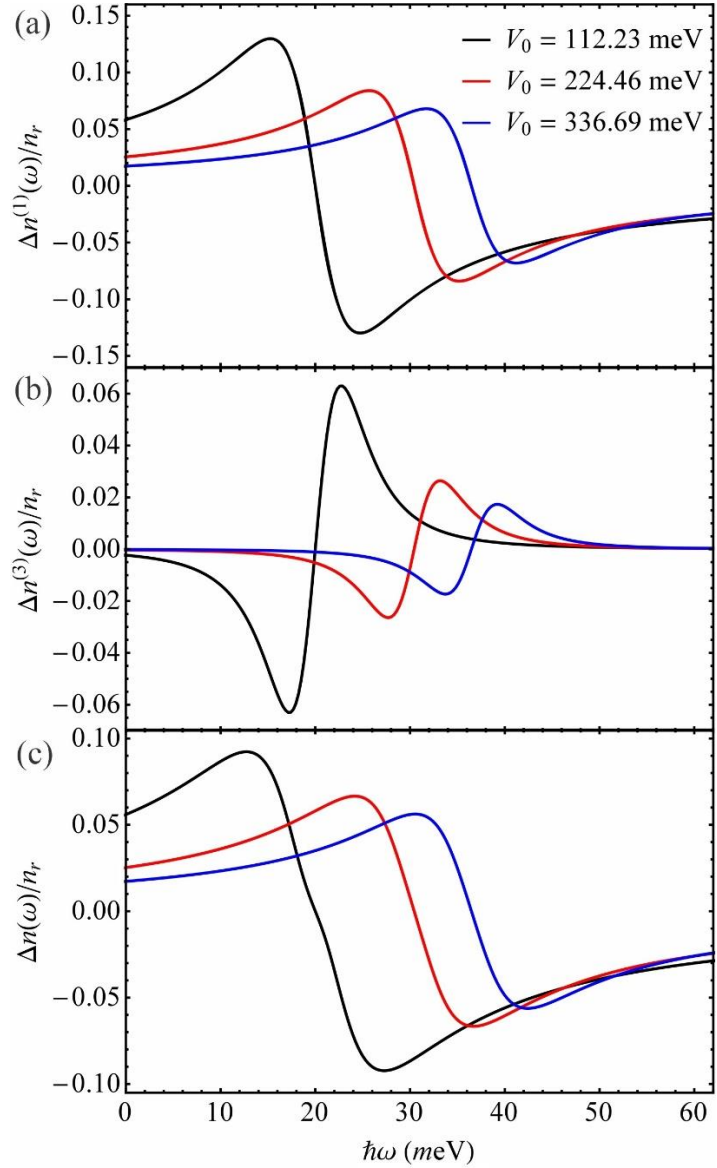

Fig. 7. The (a) linear $\Delta n^{(1)}(\omega)/n_r$, (b) third-order nonlinear $\Delta n^{(3)}(\omega)/n_r$ and (c) total $\Delta n(\omega)/n_r$ relative changes in the refractive index as a function of the incident photon energy $\hbar\omega$ for three different values of the potential height $V_0 = 112.23, 224.46, 336.69\,\text{meV}$ with dot radius $R_0 = 4\,\text{nm}$ and optical intensity $I = 8\times10^8\,\text{W/m}^2$.

The linear, third-order nonlinear and total refractive index changes are plotted as a function of incident photon energy $\hbar\omega$ for different values of the barrier height $V_0$ in Fig. 7. The radius of QD and the incident optical intensity are $R_0 = 4\,\text{nm}$ and $I = 8\times10^8\,\text{W/m}^2$. It can be seen from the figure that with increasing the barrier height $V_0$ the linear, third-order nonlinear and total relative changes in refractive index decrease and their position shift to the higher energies (blue shift). The shift of peaks is due to the fact that the increasing potential height $V_0$ results in the energy difference between the ground and excited states will increase. The physical origin of the peaks reduction with the increase of potential height $V_0$ is because the matrix element $M_{21}$ decreases with increasing $V_0$.

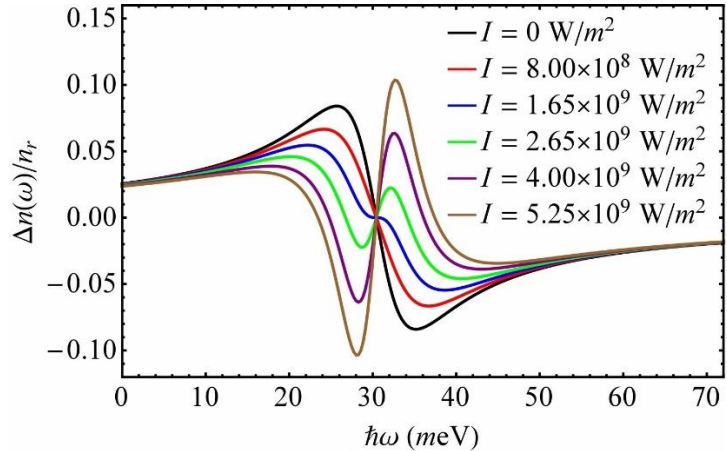

Fig. 8. The total relative $\Delta n(\omega)/n_r$ changes in the refractive index as a function of the incident photon energy $\hbar\omega$ for six different values of the incident optical intensity $I = 0,\ 8\times10^8, 1.65\times10^9, 2.65\times10^9, 4\times10^9, 5.25\times10^9\,\text{W/m}^2$ with dot radius $R_0 = 4\,\text{nm}$ and potential height $V_0 = 224.46\,\text{meV}$.

Fig. 8 shows the total relative changes in refractive index as a function of the incident photon energy $\hbar\omega$ for six different values of the incident optical intensity $I$. The dot radius and potential height are fixed as $R_0 = 4\,\text{nm}$ and $V_0 = 224.46\,\text{meV}$, respectively. As it can be observed from Eqs. (36)-(38), the linear relative change in refractive index does not depend on the optical intensity, thus, the optical intensity dependence of the total relative change in refractive index is determined via the third-order nonlinear term which linearly depends on *I*. As the intensity increases, the total refractive index change alters, together with the magnitude and sign. The linear refractive index change dominates at lower intensities and by increasing the intensity the contribution of the third-order nonlinear term will be dominant. Thus the shape of the total changes in refractive index will be modified accordingly.

## 4. Conclusion

In this work, the linear, third-order nonlinear and total absorption coefficients and refractive index changes have been studied theoretically for spherical quantum dots. For this purpose, the inversely quadratic Hellmann potential has been adopted. In order to find analytical expressions for the eigenvalues and wave functions, the Schrödinger equation has been solved by applying the Nikiforov-Uvarov method. By varying the parameters, i.e., dot radius and potential height in the model, the peak positions of absorption coefficients and refractive index changes can be shifted to lower or higher energies. In addition, the incident optical intensity does not change the position of resonant point of the total absorption coefficients and refractive index changes, only has a strong impact on their amplitudes.

**Acknowledgements**

The authors thank the Referees for a thorough reading of our manuscript and for constructive suggestions. L. M. and L.-M. P. Ț. acknowledge the financial support from a grant of the Romanian Ministry of Education and Research, PCCI UEFISCDI, project number PN-III-P1-1.2-PCCDI-2017-0338/79PCCDI/2018, within PNCDI III. L.-M. P. Ț. also acknowledges the support from the Romanian Ministry of Research and Innovation through Program 1 - Development of the national research and development system, subprogram - Institutional performance with project number 32PFE/19.10.2018 for a mobility program which made this collaboration possible. C. P. O. acknowledges the support from ICTP for a mobility program which led to this collaboration. A.-A. Farcaș acknowledges the financial support from the Romanian Ministry of Education and Research, Nucleu-Program, project PN19 35 02 01.